\documentclass[10pt,twocolumn]{article}
\usepackage[top=2cm, bottom=2cm, left=1.8cm, right=1.8cm]{geometry}
\usepackage{times}  %
\usepackage[hyphens]{url}
\usepackage{graphicx}  %
\usepackage[keeplastbox]{flushend}
\usepackage[hyphens]{url}  %
\usepackage{graphicx} %

\usepackage{url}                                %
\usepackage{array,multirow,graphicx,adjustbox}  %
\usepackage{booktabs}                           %
\usepackage[utf8]{inputenc}
\usepackage[ruled,algosection,noend,linesnumbered]{algorithm2e}
\usepackage{float}
\usepackage{paralist}
\usepackage{amsmath}
\usepackage{amsfonts}
\usepackage{xcolor}
\usepackage{csquotes} 
\usepackage{tabularx}
\usepackage{makecell}
\usepackage{pbox}
\usepackage[hang,flushmargin]{footmisc}
\usepackage{flushend}
\usepackage{xspace}
\usepackage{multicol, lipsum}
\usepackage[T1]{fontenc}

\usepackage{xcolor}
\usepackage{booktabs}
\usepackage{graphicx}
\usepackage{paralist}
\usepackage[small,bf]{caption}
\usepackage{subcaption}
\usepackage[hang,flushmargin]{footmisc}

\usepackage[compact]{titlesec}
\titlespacing*{\section}{0pt}{*4}{4pt}
\titlespacing*{\subsection}{0pt}{*3}{3pt}
\usepackage{xspace}

\makeatletter
\def\url@leostyle{%
  \@ifundefined{selectfont}{\def\UrlFont{}}%
  {\def\UrlFont{}}%
}
\makeatother
\usepackage[hyphenbreaks]{breakurl}

\usepackage[bookmarks=true, bookmarksnumbered=true, colorlinks=true, linkcolor=linkcol, citecolor=citecol, urlcolor=urlcol, hypertexnames=true]{hyperref}

\definecolor{darkgreen}{RGB}{0, 100, 0}
\definecolor{linkcol}{rgb}{0.3,0,0}
\definecolor{citecol}{rgb}{0.3,0,0}
\definecolor{urlcol}{rgb}{0.3,0,0}

\makeatletter
\def\url@leostyle{%
  \@ifundefined{selectfont}{\def\UrlFont{\small}}%
  {\def\UrlFont{}}%
}
\makeatother
\newcommand{\descr}[1]{\smallskip\noindent\textbf{#1}}

\let\OLDthebibliography\thebibliography
\renewcommand\thebibliography[1]{
  \OLDthebibliography{#1}
  \setlength{\parskip}{0pt}
  \setlength{\itemsep}{1pt plus 0.2ex}
}

\begin{document}

\pagestyle{plain}

\title{\bf On the Feasibility of Acoustic Attacks Using\\Commodity Smart Devices}

\author{Matt Wixey, Shane Johnson, and Emiliano De Cristofaro\\
{\normalsize University College London}\\[-0.5ex]
{\normalsize \{matthew.wixey.17,shane.johnson@ucl.ac.uk,e.decristofaro@ucl.ac.uk\}}}

\date{}

\maketitle

\begin{abstract}
Sound at frequencies above (ultrasonic) or below (infrasonic) the range of human hearing can, in some settings, cause adverse physiological and psychological effects to individuals.  In this paper, we investigate the feasibility of cyber-attacks that could make smart consumer devices produce possibly imperceptible sound at both high (17--21kHz) and low (60--100Hz) frequencies, at the maximum available volume setting, potentially turning them into acoustic cyber-weapons. To do so, we deploy attacks targeting different smart devices and take sound measurements in an anechoic chamber. For comparison, we also test possible attacks on traditional devices.

Overall, we find that many of the devices tested are capable of reproducing frequencies within both high and low ranges, at levels exceeding those recommended in published guidelines. Generally speaking, such attacks are often trivial to develop and in many cases could be added to existing malware payloads, as they may be attractive to adversaries with specific motivations or targets. Finally, we suggest a number of countermeasures, both platform-specific and generic ones. 
\end{abstract}

\section{Introduction}
\label{sec:intro}
Concerns about the potential for malware to harm citizens by compromising smart consumer devices have become increasingly prevalent~\cite{krotofil2014,sturm2014,chen2014}. 
However, while previous research has extensively focused on ``traditional'' IoT malware (e.g.,~exfiltrating confidential information~\cite{yang2017survey}, mounting Distributed Denial of Service (DDoS) attacks~\cite{kolias2017ddos}, etc.), less attention has been paid to the ability to {\em directly} cause material harm to users of compromised systems. 

In this paper, we focus on malware that may have direct {\em psychological} and/or {\em physical} impacts on the users of hosts under attack. 
This is different from malicious code indirectly causing harm, e.g., by disrupting the power grid or a water treatment plant.
In particular, we set out to study the feasibility of attacks developed to control consumer devices and make them reproduce {\em sound} that is likely to be {\em imperceptible} to a significant proportion of the population at both high (17kHz--21kHz) and low (60Hz--100Hz) frequencies, but also possibly damaging to them. 
If such noise is emitted at sufficient levels, and for sufficient periods of time, a number of short-term and long-term adverse physical and psychological effects may occur, affecting users of that device as well as those nearby (see Section~\ref{sec:background} for a primer on high- and low-frequency noise and their adverse affects). 

We rely on an experimental methodology to design and deploy a range of attacks on a variety of devices and take sound measurements in an anechoic chamber in order to assess the capabilities of a sample of consumer equipment both in terms of the frequencies and sound levels achievable. 
We first report on a few {\em smart} devices, whereby smart here denotes  devices with a remote or local network interface, including Internet-connected speakers and headphones. 
For comparison, we also run attacks against more ``traditional'' devices -- namely, parametric and vibration speakers, a loudspeaker, and a vehicle-mounted PA system -- which rely on intended control channels like Bluetooth, or on physical access to the device in question.

Overall, we show that we can indeed re-purpose some devices for local or remote acoustic attacks by an attacker with the objective of causing direct harm to humans. %
One of our attacks depends on a known vulnerability in a specific device, while the others use customized malware relying on common functionalities to manipulate system volumes and taking advantage of the ability of consumer audio equipment to reproduce sound at low and high frequencies. 
Of the ten device set-ups tested, four (two smart, two traditional) were capable of emitting high-frequency noise (HFN) and/or low-frequency noise (LFN), at levels which exceed published guidelines relating to the maximum recommended levels.
More specifically, a smart speaker and a headphones set did that for both HFN and LFN; a parametric speaker for HFN only; and a loudspeaker for LFN only. 

To the best of our knowledge, ours is the first attempt to demonstrate the feasibility of extending malware and cyber-attacks into the field of acoustic weapons.
Also note that all the devices we experiment with are publicly available, relatively modern and inexpensive, and commonly purchased in both home and business contexts.
For safety reasons, we do not provide full details of these devices.

We also show that attacks such as these can, in some cases, have unintended but significant effects on the physical equipment itself; we were able to cause permanent damage to the smart speaker, preventing it from reproducing frequencies above 5kHz, by playing a particular frequency for a few minutes at maximum volume. 
This was disclosed to the manufacturer, who subsequently notified us that a mitigation would be applied to address this issue.
Finally, we discuss a number of possible countermeasures, including an open-source application which can be used to detect exposure to audio at specified frequencies.
\section{Background}
\label{sec:background}
\descr{HFN and LFN.} Frequencies believed to be above or below the range of human hearing are often defined as {\em ultrasonic} or {\em infrasonic}, respectively.
More specifically, the former encompass higher frequencies, usually 20kHz and higher~\cite{howard2005}, while the latter are in the range 0--20Hz~\cite{leventhall2003}.
However, as highlighted by Duck and Leighton~\cite{duckleighton2018}, founding a definition on a lack of a property (namely, non-audibility) is problematic, particularly with a concept that is highly subjective.
In this paper, we focus on High-Frequency Noise (HFN) in the 17kHz--21kHz range, due to the reported capacity of some consumer devices, such as mobile phones, to produce noise at approximately these frequencies~\cite{filonenko2010,leighton2018}, as well as Low-Frequency Noise (LFN), typically described as 20-- 200Hz~\cite{cocchi1992,bolin2011, bengtsson2003,storm2009}.
However, for the latter, we restrict testing to the 60--100Hz range, following the results of a pilot study, presented in ~\ref{sec:pilot}, which indicated that available devices would not be capable of reproducing lower frequencies.

\descr{Hearing Thresholds.} A common misconception is that healthy humans are unable to perceive noise above a 20kHz threshold or below 20Hz~\cite{durrant1995}. 
However, perceptibility does not solely depend on theoretically defined cut-off points. %
In fact, the mechanisms of perception of both low and high frequencies are complex and not fully understood~\cite{koch2017}, and, %
there is a significant amount of variation in the ability of people to detect HFN and LFN~\cite{wieringen2018,leighton2018,leventhall2003}.
For instance, some individuals have reportedly been able to hear frequencies above 17.8kHz~\cite{duckleighton2018} or higher~\cite{dolder2018,koch2017,oohashi2000}, or down to 1.5Hz in certain conditions~\cite{leventhall2009}. 
To a large extent, this %
depends on a number of factors, including the sound pressure level (SPL), levels of background noise, etc. 
Additionally, lower frequencies may be perceived, but not necessarily as ``sound''~\cite{leventhall2003}, and the generation of high frequencies may cause subharmonics in the audible range~\cite{howard2005,duckleighton2018, ashihara2006}. 
However, there is a general consensus that the likelihood that people can hear sounds declines non-linearly with increasingly higher~\cite{ashihara2006} and lower frequencies~\cite{muhlhans2017} and that, for the former, hearing thresholds generally increase with age~\cite{wieringen2018, macca2015}. Put simply, it is likely that many people, particularly older adults, cannot hear sound at the ranges we test in this paper. 

\descr{Adverse Effects of HFN/LFN.} HFN and LFN have both been associated with adverse physiological and psychological effects.  
However, as with perceptibility, susceptibility is again likely to differ significantly between individuals~\cite{leighton2016,qibai2004}. 
While there have been no reports of high frequencies causing permanent hearing loss~\cite{howard2005}, there have been numerous reported cases of ultrasound having adverse effects on hearing~\cite{duckleighton2018}, including temporary threshold shifts~\cite{acton1967}; reductions in hearing sensitivity in the audible range~\cite{macca2015, chopra2016, wilson2002,grzesik1986}; neurasthenia, cardiac neurosis, hypotension, bradycardia, and functional changes in cardiovascular and central nervous systems~\cite{smagowska2013}. 
Permanent threshold shifts have only been associated with high frequency exposure in the presence of high levels of lower frequencies~\cite{leighton2018}.
High frequencies have also been linked to more subjective effects, including nausea, fatigue, and headaches~\cite{duckleighton2018,vongierke1992,howard2005}; tinnitus and ear pain~\cite{chopra2016, fletcher2018part1}; irritation~\cite{ueda2014}; somnolence, dizziness, palpitations, and decreased concentration~\cite{smagowska2013}.

Although LFN has been associated with temporary threshold shifts~\cite{leventhall2003}, and some correlation observed with various conditions such as heart ailments, chronic insomnia~\cite{mirowska2000}, and elevated levels of cortisol~\cite{bengtsson2003}, annoyance is often the most common response~\cite{stansfeld2015,persson1988,pawlaczyk2005,storm2009}.
Other subjective effects include headaches and palpitations~\cite{moller2002}; deterioration in task performance~\cite{benignus1975,bengtsson2003}; decreased productivity~\cite{kaczmarska2007}; and lower levels of cooperation and agreeableness~\cite{benton1997}. 
These subjective effects are often reported even at relatively moderate levels of between 40 and 45 dB(A)~\cite{bengtsson2003,benton1997,pawlaczyk2005,perssonbjorkman1988}, with noise sensitivity reported to be a consistent predictor of depressive symptoms and psychological distress~\cite{stansfeld2015}.

\descr{\em Remarks.} It is crucial to highlight that there are often issues with definitively establishing a causal relationship between HFN and LFN and adverse effects. 
Data is often sparse and anecdotal~\cite{leighton2018}, and %
detailed knowledge of the ``noise dose'' -- including both the level and the duration of the exposure -- is required in order to evaluate effects~\cite{andringa2013,dolder2018}. 
This data is often unavailable, and has not been explored in detail~\cite{lawton2001}. 
In fact, many effects have not been successfully reproduced in laboratory settings~\cite{fletcher2018effects2}, although this may be in part due to ethical restrictions on exposing human subjects to potentially dangerous SPLs~\cite{leighton2018,fletcher2018part1,fletcher2018effects2}, or to the possibility of nocebo effects~\cite{fletcher2018effects2}.  
However, as pointed out by Leighton~\cite{leighton2018}, %
while it is not possible to make definitive statements about causality, there exists a significant evidence base for the threat of adverse effects at lower intensities in a subset of the population. Moreover, these threats are sufficiently evidenced that a number of organizations and researchers have developed guidelines detailing recommended maximum permissible exposure levels for both HFN and LFN.

\descr{Exposure Guidelines.} Most exposure guidelines, particularly for for HFN, have been developed for occupational use~\cite{leighton2018}. %
There are often significant differences in the way these levels are calculated and implemented, and in the proposed recommendations for comparison and evaluation against them. 
In this paper, we will not assess the merits, or lack thereof, of individual guidelines, but will instead use them to compare our generated levels.

Leighton~\cite{leighton2016} presents a compendium of maximum permissible sound pressure levels (MPSPLs), the means and medians of which are reported in Table~\ref{tab:leighton_table}. 
As noted in Leighton's follow-up work~\cite{leighton2018}, many of these guidelines are based on small samples, often only including adult males and predominantly focusing on occupational environments rather than public exposure. 
Although the research base may be too small to support such guidelines~\cite{leighton2016}, there is, at least, something of a consensus~\cite{howard2005}, particularly for the fact that A-weighting -- commonly used for exposure guidelines in the audible range -- is limited, as it significantly underestimates higher frequencies~\cite{lawton2001,leighton2018}. Our measurements are taken using Z-weighting -- a flat frequency response 10Hz--20kHz, which, unlike A-weighting, does not apply any attenuation for sounds above or below the commonly understood ``audible range.''
In Section~\ref{sec:results}, we will compare to the various guidelines using this weighting, with notation LZ\textsubscript{eq}.

\begin{table}[tbp]
\setlength{\tabcolsep}{2.25pt}
\resizebox{\linewidth}{!}{
\begin{tabular}{lrrrrrrrrr}
\toprule
\multicolumn{2}{r}{\textbf{(kHz) ~ 8}}  & \textbf{10} & \textbf{12.5} & \textbf{16} & \textbf{20} & \textbf{25} & \textbf{31.5} & \textbf{40} & \textbf{50} \\ 
 \midrule
Mean & 80.00 & 83.08 & 82.67 & 83.89 & 96.91 & 111.08 & 113.91 & 114.09 & 115.28 \\
Median & 80.00 & 80.00 & 80.00 & 80.00 & 105.00 & 110.00 & 110.00 & 110.00 & 110.00 \\ \bottomrule
\end{tabular}
}
\vspace{-0.15cm}
\caption{Mean and median of maximum permissible sound pressure levels (MPSPLs), for different frequencies, as per~\cite{leighton2016}.}
\label{tab:leighton_table}
\vspace{-0.3cm}
\end{table}

We are unaware of any similar compendium relating to safety guidelines regarding exposure to LFN. 
However, several bodies have published reference curves for the assessment of {\em disturbance} caused by LFN.
Unfortunately, the methods used to calculate these curves differ significantly. 
For our analysis, we use a reference curve proposed in 2011~\cite{moorhouse2011}, reported in Table~\ref{tab:moorhouse_table}, which was devised after an assessment of previously published reference curves.
This reference curve proposes the use of L\textsubscript{eq}. 
Since no weighting is applied, we measure and compare our results using LA\textsubscript{eq} in third-octave bands (TOBs). There is a general consensus that A-weighting may underestimate the effects of LFN~\cite{leventhall2003,vasudevan1982,kjellberg1984} due to its attenuation at lower frequencies, %
so this should be taken into account when reviewing our results in Section~\ref{sec:results}.

\section{Related work}
\label{sec:relatedwork}

\descr{High-Frequency Noise (HFN).} Previous work has studied HFN in the context of enabling or supplementing attacks. %
More specifically, researchers have used ultrasound to create covert communication channels~\cite{hanspach2014,wixey2017,deshotels2014}, finding that many consumer devices are capable of emitting HFN~\cite{leighton2018,filonenko2010}. 
Other research involving HFN includes the disruption of obstacle-detection systems by introducing attacker-controlled ultrasound to perform echolocation jamming~\cite{yan2016,wixey2017}. 
Also, Bolton et al.~\cite{bolton2018} explore the capability of both audible and inaudible noise to corrupt data being written to hard disk drives, while Mavroudis et al.~\cite{mavroudis2017} first and Cunche et al.~\cite{cunche2018} later investigate the use of ultrasonic beacons as tracking devices in the context of targeted marketing, exploring related privacy implications. 
Finally, we are not aware of any security-related research into the use of low frequencies.

\descr{Physical Harm.}
Researchers have examined the ways in which malware could be used to cause physical harm in a number of contexts, e.g., embedded medical devices~\cite{halperin2008pacemakers,williams2015cybersecurity,rushanan2014sok}. Depending on the specific device and context, an attacker can cause significant, life-threatening harm by exploiting vulnerabilities in such systems. Other research in a similar vein has explored the physical risks posed by vulnerabilities in transport systems, such as connected cars~\cite{ben2013towards} or air traffic control systems~\cite{brody2018malware}, as well as the manipulation of IoT devices to force them to strike humans~\cite{rios2017iot}.

Overall, there has been little research on the ability of attackers to {\em directly} harm users through malware and other attacks, i.e., by manipulating the ordinary outputs of devices to cause adverse effects. 
One exception appears to be work on the inducement of epileptic seizures. 
Poulsen~\cite{poulsen2008} reports on a series of attacks against a forum for epilepsy sufferers: attackers uploaded flashing images, successfully causing a number of seizures in forum users. 
Oluwafemi et al.~\cite{oluwafemi2013} and Ronen and Shamir~\cite{ronen2016} also discuss vulnerabilities in connected lighting devices, finding that an attacker can cause vulnerable systems to flash in patterns consistent with those known to induce seizures. 
To the best of our knowledge, our work is the first to examine the feasibility of acoustic attacks using malware or cyber-attacks.

\descr{Acoustic Weapons.} Perhaps as a result of a substantial, albeit often anecdotal, evidence base, %
there has been significant historical interest in the development of devices that could be used to deliberately expose people to harmful levels of sound.
This topic has been the subject of frequent misunderstandings and rumors~\cite{vinokur2004,muhlhans2017}, and while it is generally agreed that, in principle, acoustic weapons could be used to covertly generate adverse effects in humans~\cite{leighton2018,arkin1997}, there would be significant practical problems in deploying such devices, which serves as motivation to our work.

Altmann~\cite{altmann2001} notes that threshold shifts, not being immediately felt or causing an immediate impact, would be of little interest to those deploying acoustic weapons, and that it would be challenging to cause targeted, directional effects. 
Bartholomew and Perez ~\cite{bartholomew2018} agree with the latter point, arguing that the need for close proximity, the required size of the acoustic weapon, and the rapid diffusion of ultrasound, would make such weapons impractical. 
However, as our results suggest, the deployment of acoustic attacks in the context of cyber-attacks could to some extent negate these disadvantages. %
Attackers may be able to affect victims over extended periods of time, particularly as users of consumer devices are typically within fairly close proximity to them, often for long periods. Therefore, concerns over practicality with regards to size and diffusion would seem less relevant with the advent of smart devices.

\descr{\em Remarks.} 
Overall, while previous work has explored the ability of cyber-attacks to cause physical or psychological harm to users, %
there has not yet, to the best of our knowledge, been any empirical work on the capacity of malware to create localized acoustic weapons.

\begin{table}[t]
\centering
\setlength{\tabcolsep}{2.5pt}
\resizebox{0.9\linewidth}{!}{
\begin{tabular}{rrrrrrrrrrrrr}
\toprule
\textbf{(Hz) ~10} & \textbf{12.5} & \textbf{16} & \textbf{20} & \textbf{25} & \textbf{31.5} & \textbf{40} & \textbf{50} & \textbf{63} & \textbf{80} & \textbf{100} & \textbf{125} & \textbf{160} \\ \midrule
92 & 87 & 83 & 74 & 64 & 56 & 49 & 43 & 42 & 40 & 38 & 36 & 34 \\ \bottomrule
\end{tabular}
}
\vspace{-0.15cm}
\caption{Reference curve by Moorhouse et al.~\cite{moorhouse2011} for assessing LFN. Levels shown as L\textsubscript{eq} in centered third-octave bands (TOBs).}
\label{tab:moorhouse_table}
\vspace{-0.3cm}
\end{table}

\section{Methodology}
\label{sec:method}
We now present our methodology to assess the feasibility of acoustic attacks on commodity hardware.
We do so on several commonly purchased and publicly available ``smart'' devices that can produce sound, namely: laptops, mobile phones, and smart speakers. We also include a pair of smart headphones in this category.
(Overall, {\em smart} here denotes devices with a remote or local network interface, including Internet-connected speakers and headphones.)
As a comparison, we also use more traditional audio equipment: parametric speakers, loudspeakers, vibration speakers, and a vehicle-mounted PA system. 

As detailed in this section, our methodology involves: (1) designing and deploying attacks to target each device in our testbed, (2) forcing them to play pre-prepared audio of selected low and high frequency tones, and (3) measuring the results using sound level meters in an anechoic chamber.

\subsection{Pilot study}
\label{sec:pilot}
In order to obtain an initial indication as to whether consumer devices were indeed capable of producing HFN and LFN, we conducted a  pilot study using four of the selected devices: a laptop, mobile phone, loudspeaker, and smart speaker. 
The experiments were also conducted in the anechoic chamber used for our full study, using the same proof-of-concept attacks (presented in Section~\ref{sec:attacks}). %
As the goal was not to precisely measure audio emissions, but to simply assess whether the devices could reproduce the required frequencies, we used two publicly available Android apps, Ultrasound Detector~\cite{ultradetector} and Infrasound Detector~\cite{infradetector},
and a factory-calibrated Dayton Audio iMM-6 external microphone connected to an Android phone. 
This is reasonable as modern smartphones are generally considered suitable for occupational noise measurements, within the limitations of the device in question~\cite{kardous2014}.

Our findings showed that, while all the devices appeared to be capable of reproducing HFN, from around 60.5dBSPL to 91.5dBSPL, only the smart speaker and loudspeaker were capable of reproducing LFN at a reasonable level (50Hz at 63.4dBSPL).

We also observed a distinct increase in temperature in the smart speaker, following the production of HFN at maximum volume. 
More specifically, the speaker became noticeably hot to the touch and gave off a strong odor of burnt plastic after the HFN testing runs. 
However, we did not observe any smoke or flame coming from the device, and assumed that the production of HFN at maximum volume had caused some form of internal damage to an electronic component. As a result of this observation, we opted to include heat measurements in our full study.

Moreover, some time after the pilot study, we noticed that the speaker's ability to reproduce higher frequencies had been impaired. 

\subsection{Experimental Setup}
\label{sec:experimentalsetting}

\descr{Testing Environment.} Our experiments were conducted in an anechoic chamber at our institution. While this was necessary in order to accurately and safely measure emitted noise, it should be noted that in a real-world environment, ambient sounds and certain types of environment may amplify or reduce the effects of LFN or HFN.
Owing to the nature of the study, and the reported association between high levels of LFN/HFN and adverse effects on people, we did not use human subjects for this research; instead, we measured the sound emitted from each device as a consequence of the attacks, and assessed whether or not the resulting levels exceeded published maximum permissible levels.

\descr{Ethics.} A full risk assessment was conducted prior to the experiment, and ethics approval was obtained from our institution. 

\descr{Device Set-Ups.} %
Our experiments involved: 1) a Windows laptop, 2) an Android smartphone, 3) a pair of wireless over-ear headphones, 4) a smart speaker, 5) a loudspeaker, 6) a vibration speaker, 7) a parametric speaker, and 8) a vehicle-mounted PA system. 
To minimize risks to the general public, we do not include details of specific brands and models, or the code for our attacks.

In addition to the attacks developed for each device, we also expanded our testbed to include 9) three Windows laptops and 10) three Android smartphones, which were attacked at the same time during the test. 
We included these additional tests to assess the scalability of the attacks, given that in some environments, e.g., open-plan offices, several such devices might realistically be in close proximity to each other. 
In that case, the resulting levels may increase~\cite{hansen2001fundamentals} and thus their effects as well. 

\descr{Procedure.}
We placed each device inside an anechoic chamber, along with a Class I sound level meter, spot-calibrated by the supplier, and placed at a distance of one meter from the device. 
For the HFN tests, we used a Svantek 977A sound level meter with GRAS 40AM microphone, and a Svantek 979 sound level meter with GRAS 40AE microphone for the LFN tests.
Each device was made to play or stream a WAV audio file, generated online\footnote{\url{https://www.audiocheck.net}}, with a sine wave tone at a sample rate of 44.1kHz %
at a single frequency. 
We initiated each tone on each device for a period of ten minutes, while the chamber remained shut, using a specific attack developed to test that particular device, as discussed later in Section~\ref{sec:attacks}. 
Following each ten-minute period, the anechoic chamber was opened and readings were taken from the sound level meter.

We also measured the surface temperature of each device using an infrared thermometer, before and after each testing period, to assess whether the production of LFN or HFN caused an increase in temperature. 
We report these measurements in Section~\ref{sec:temp}.

\descr{Frequency Measurements.} Note that all but one of the frequencies being tested was below 20kHz, thus, we took measurements using Z weighting (a flat frequency response in the band 10Hz--20kHz) in these cases. 
For test runs involving the ultrasonic frequency (21kHz), we used a proprietary high-pass filter weighting developed by the sound level meter manufacturer, known as HPE (high-pass extended). 
For test runs involving LFN, our original intention was to use G-weighting, which is the ISO 7196:1995 standard for measuring infrasound in the band 1Hz--20Hz. 
However, the results of our pilot study indicated that many consumer devices were not capable of producing noise in this range. 
Therefore, we increased the frequencies being tested to 60Hz, 80Hz, and 100Hz. These still fall within most definitions of LFN and are still associated with reported adverse effects, as discussed in Section~\ref{sec:relatedwork}, but are not infrasonic, and were thus suitable for Z-weighted measurements rather than G-weighting, which is designed exclusively for infrasound~\cite{koch2017}.

\subsection{Attacks on Smart Devices}
\label{sec:attacks}

\descr{Smart Speaker.}
Our attack against the smart speaker relied on a (previously disclosed) vulnerability affecting a number of smart audio products;
specifically, that no authentication is required between the smart speaker and the controller.
That is, they simply need to be on the same local network. 
As previously discussed, we do not disclose details of specific models affected for safety reasons, however, we can say that our experiments are performed on a speaker released a couple of years ago for around \$200.

In theory, exploiting this vulnerability requires an attacker to be joined and authenticated to the user's WiFi network, and therefore nearby.  
However, %
a number of these smart speakers are exposed on the Internet, even though remote access is not the default configuration of the speakers.
Indeed, a query on the search engine Shodan for the brand name and port associated with the smart speaker showed that over 5,000 results in five countries, as of Spring 2019, meaning that an attacker could find and remotely control these devices without authentication, including launching our attack.

To execute that, we wrote a Python script which scans the current local network for smart speakers of a particular brand. 
If any are found, and are inactive, the script retrieves the current volume level as an integer and stores it as a variable, raises the volume to maximum, and streams a requested WAV file hosted on a web server controlled by the attacker.

\descr{Headphones.} We also used wireless headphones (released approximately two years ago, costing around \$400). Note that we did not attack the headphones directly, but tested the capability of the headphones to reproduce HFN and LFN using the Windows malware described above, by connecting the headphones to the laptop over Bluetooth. Whilst some of the ``traditional'' devices we test also use Bluetooth, headphones are reported to be increasingly attached to smart devices~\cite{vaidya2018evaluation,mohammadpoorasl2018prevalence} and so we include them in the ``smart'' category, as attacks using headphones are not reliant on attacking an intended controlled channel such as Bluetooth, but could be achieved by attacking a smart device to which they may be attached.
Unlike the other tests, where the sound level sensors were placed one meter away from the device, here we placed them approximately one centimeter from the headphone's speakers, aiming to simulate as closely as possible the effect a user would experience while wearing the device. 

\descr{Windows Laptop.} We developed proof-of-concept Windows malware, with WAV files corresponding to each target frequency embedded in the malware as resource files. 
The malware contacts a simple command-and-control server to retrieve commands. If the command to play a specific frequency is received, the malware will store the current volume level as an integer, increase the volume to maximum, and play the requested tone. Upon receiving the 'stop' command, the malware will restore the volume level to its previous value.
For our extended test using three laptops, we infected each machine with the same proof-of-concept malware and controlled all the infected hosts simultaneously using the same C2 server. 
Note that we experiment on mid-range laptops released a couple of years ago, priced in the order of \$1,000.%

\descr{Android Phone.}
We also developed a proof-of-concept Android app to simulate a malware-infected phone, with WAV files corresponding to each target frequency embedded in the malware as resource files.
This app has the same functionality as that described for the Windows laptop malware, and our extended test using three phones similarly employed simultaneous remote control using one C2 server. 
Again, we used mid-range phones released about two years ago, priced at around \$200.

\subsection{Attacks on Traditional Devices}

\descr{Vibration Speaker and Loudspeaker.}
Vibration speakers differ from traditional speakers in that they do not use a diaphragm cone. 
Instead, the speaker's coil is fixed to a movable plate, which pushes against the surface the speaker is placed on, causing that surface to vibrate and emit sound. 
Due to the lack of a diaphragm, these speakers typically have a smaller profile %
and can be attached to a variety of surfaces unobtrusively, possibly making them an attractive choice as repurposed acoustic weapons -- either through an attacker executing an attack against another user's device, or purchasing and using their own.

The vibration speaker we used was controlled through Bluetooth, as was the loudspeaker. 
For both of these devices, we paired the speaker to the Android phone and used our Android malware to play the targeted tones through these speakers.
The loudspeaker model is about two years old and costs around \$50, while the vibration speaker model is five years old and cost around \$70.

\descr{Parametric Speaker.} Parametric speakers use ultrasonic carrier waves, typically at 40kHz, to transmit high-intensity directional audio in a relatively small area of focus, essentially creating a ``beam'' of sound (please refer to~\cite{pompei2002sound} for further details on the operation of parametric arrays).
These devices are often used to transmit focused advertisements or informational messages at trade shows or retail stores, or to create the (auditory) illusion of sound appearing to come from another source.

Note that the speaker we used has no smart capabilities and no remote or local command channels; instead, a standard 3.5mm audio cable is used to connect the speaker to an audio source. 
For our tests, we connected this speaker to our Windows laptop and used the Windows malware to play the targeted tones through the speaker. As this speaker is known to use 40kHz carrier waves, we also measured its emissions at this frequency using the HPE filter.
This device is roughly the size of a mobile phone, and available for purchase online at a moderate cost, around \$250, therefore, it could be used as a low-cost portable acoustic weapon by an attacker -- particularly as the directional nature of the transmitted audio may allow them to target a specific location.

\descr{PA System.}
Finally, we used a vehicle-mounted PA system, which, like the parametric speaker, has no network interfaces. 
It automatically plays audio upon inserting a storage device, e.g., a USB drive or a SD card. 
For each test, we placed an audio file %
on a USB drive that was plugged into the device. 
So that we could exit the anechoic chamber safely, the recording had a thirty second delay before the ten minutes of a given tone were played.

An attacker seeking to attack this particular system would therefore need to have physical access to the device, have a pre-prepared storage device containing audio in the required format, and be able to insert that storage device into the system without arousing suspicion.  
Alternatively, as with the parametric speaker, the attacker could purchase a similar device with the intention of using it as a `mobile' acoustic weapon when mounted on a vehicle.

\descr{Additional attacks.}
\label{sec:additional_attacks}
We devised two more possible attacks in addition to those described in Section~\ref{sec:attacks}, which, rather than targeting specific devices, would be suitable for deployment generally. However, as these would have utilized the same audio components being tested, and since they rely on targeted users having their volume set high enough to cause harm, we did not include them in our testing plan. 
Nevertheless, they might remain plausible attack scenarios, thus, we briefly discuss them here.
The first additional attack relies on the HTML5 audio tag; specifically, the autoplay attribute. In this instance, an attacker would need to persuade a victim to visit a particular attacker-controlled server, and a selected tone hosted on the attacker's server would autoplay at whatever volume is currently set, without the user's knowledge---even though, depending on the browser being used, a small speaker icon might appear on the relevant tab. As it is not possible for code on a webpage to manipulate a user's system volume, the efficacy of this attack, in terms of causing harmful levels of audio, depends on the volume set on the user's device.

Another attack involves the deliberate manipulation and insertion of particular audio into a pre-existing audio track. 
Here the attacker may have access to a legitimate audio file that they know an intended victim will play at some point. This could be, for instance, a YouTube video, a film soundtrack, or some other audio. Using an audio editor, the attacker could decrease the level of the legitimate audio, and insert an ultrasonic or low-frequency tone of their choosing at a much higher level. 
Upon playing the manipulated file, the user is likely to assume that they do not have their system volume turned up high enough, or that the legitimate audio was not recorded at sufficient levels, and as a result may significantly increase their system volume -- leading to exposure to potentially harmful levels of the attacker-selected tone. As with the previous attack, this approach would require the system volume of the device in question to be high enough to emit harmful levels of audio.

\descr{\em Remarks.} Our primary interest in developing these attacks was to assess whether the proliferation of smart devices could provide a new avenue for potentially harmful acoustic attacks. 
Overall, our attacks are realistically viable in the wild and potentially very harmful.
In addition to many of the smart devices we tested being ubiquitous in a number of diverse environments, including homes, businesses, and public or social events, note that many of the attack vectors are ``generic.'' For instance, there are multiple ways to deploy malware infections on a laptop or mobile phone, and other devices, such as the headphones, could be used for attacks arising from a number of vectors. 

We also experiment with a number of traditional devices. %
These attacks are perhaps less realistic, lacking vulnerable control channels and connectivity and typically requiring either physical access or close proximity, as well the ability to pair with an unpaired Bluetooth device. However, we include them in our testing both as a comparison to the tested smart devices, and to investigate whether the abuse of more traditional consumer equipment may also be an attractive avenue for attackers.

\section{Experimental Evaluation}
\label{sec:results}
We now present the results of our experiments and the related measurements.

\descr{Overview.} Overall, we find that several devices (two smart, one traditional) were capable of producing HFN at levels exceeding many of the recommended exposure limits, and are therefore potentially able to cause temporary and permanent harm, depending on the circumstances and duration of exposure. 
Additionally, a number of devices (two smart, one traditional) were capable of producing levels at or above LFN limits.

Our experiments highlight a tendency for devices to perform better at lower high frequencies (such as 17kHz) and higher low frequencies (100Hz). 
This may present an attacker with a disadvantage if they select these frequencies, as lower HFN and higher LFN may be more likely to be perceived by users. 
An exception was the parametric speaker, which performed better at higher frequencies, presumably due to its array of ultrasonic transducers.

\subsection{High Frequency Noise}
\label{sec:hfresults}

As discussed in Section~\ref{sec:background}, we used the compendium of MPSPLs for airborne ultrasound in Leighton~\cite{leighton2016} to assess the capability of the devices to reproduce HFN. 
Results are reported in Table~\ref{tab:hfn_spl}, which show several results exceeding the mean average of these MPSPLs at relevant frequencies.

\begin{table}[t]
\centering
\small
\setlength{\tabcolsep}{3pt}
\begin{tabular}{@{}lrrrr@{}}
\toprule
& \textbf{17kHz} & \textbf{19kHz} & \textbf{21kHz(HPE)} & \textbf{40kHz} \\ \midrule
Smart speaker & \textbf{86} & 35.2 & 43.8 & - \\
Headphones & \textbf{87.5} & 81.2 & 79.8 & - \\
Laptop & 63 & 64.5 & 45.5 & - \\
Mobile phone & 59.4 & 58.3 & 16.9 & - \\
3 laptops & 65.6 & 63.8 & 57.5 & - \\
3 phones & 59.8 & 61.1 & 45.3 & - \\
Loudspeaker & 59.4 & 48.5 & 54.5 & - \\
Vehicle PA & 75.3 & 20.5 & 18.5 & - \\
Vibration speaker & 47.7 & 36.1 & 27.3 & - \\
Parametric speaker & \textbf{85.1} & 84.2 & \textbf{97.1} & \textbf{117.7} \\ \bottomrule
\end{tabular}
\vspace{-0.1cm}
\caption{Levels observed during our HFN trials. Levels (LZ\textsubscript{eq} in centered TOBs) which exceed the mean and/or median average of MPSPLs in Leighton~\cite{leighton2016}'s guidelines are in bold.}
\label{tab:hfn_spl}
\vspace{-0.15cm}
\end{table}

Note that the smart speaker produced a high of 86dB (all results LZ\textsubscript{eq}) at 17kHz, but subsequent HFN trials produced much lower levels. 
This is due to the result of internal damage caused to the speaker during the experiment, which we discuss in Section~\ref{sec:temp}. 
Taking the TOB center of 16kHz, this exceeds both the mean and median averages of the MPSLs reported by Leighton~\cite{leighton2016}.

Moreover, we found that the headphones produced a mean average of 82.83dB for the HFN trials, with a high of 87.5dB at 17kHz. 
Taking the TOB center of 16kHz, this also exceeds both the mean and median average levels.

The parametric speaker produced a mean average of 88.80dB for the HFN trials, with a high of 97.1dB at 21kHz. A high of 117.7dB was observed at 40kHz during the 21kHz trial.
The 17kHz experiment, taking a TOB center of 16kHz, exceeded the mean and median average MPSPLs, and the 21kHz result exceeded the mean average for MPSPLs at TOB center 20kHz. 
Finally, the 40kHz result (TOB center 40kHz) exceeded both the mean and median MPSPLs, indicating that the speaker's ultrasonic carrier waves at 40kHz could present a health risk. 

\subsection{Low Frequency Noise}
\label{sec:lfresults}
LFN tests generally produced lower levels than the HFN tests. 
However, as many researchers report~\cite{bengtsson2003,benton1997,pawlaczyk2005,perssonbjorkman1988}, adverse psychological effects associated with low frequency sound are often observed at relatively moderate levels. 

\begin{table}[t]
\centering
\resizebox{0.75\linewidth}{!}{
\small
\centering
\begin{tabular}{@{}lrrr@{}}
\toprule
  & \textbf{60Hz} & \textbf{80Hz} & \textbf{100Hz} \\ \midrule
Smart speaker & \textbf{47.5} & \textbf{59} & \textbf{71.6} \\
Headphones & 37.5 & 39.9 & \textbf{40.2} \\
Laptop & 2 & 0.1 & 3 \\
Mobile phone & 1 & 1.2 & 6.5 \\
3 laptops & 1.4 & -0.3 & 4.7 \\
3 phones & 3.3 & 1.6 & 12.5 \\
Loudspeaker & 38.2 & \textbf{51} & \textbf{64.2} \\
Vehicle PA & 13.7 & 22.6 & 33.7 \\
Vibration speaker & 24 & 21.1 & 18.4 \\
Parametric speaker & -0.6 & 0.5 & 28.6 \\ \bottomrule
\end{tabular}
}
\vspace{-0.1cm}
\caption{Levels observed during our LFN trials. Levels (LA\textsubscript{eq} in centered TOBs) exceeding the reference curve values are in bold.}\label{tab:lfn_spl}
\vspace{-0.4cm}

\end{table}

To compare our results to the LFN reference curve~\cite{moorhouse2011}, we apply A-weighting to the levels observed at TOB center frequencies, as shown in Table~\ref{tab:lfn_spl}. It should be noted that A-weighting results in significant attenuation at lower frequencies, down to -26.2dB in the 63Hz centered TOB (the lowest band used in our analysis), and as much as -85.4dB at 6.3Hz. As a result, the A-weighted levels are significantly lower than our Z-weighted measurements.

The loudspeaker produced a mean average of 51.13dB (all results LA\textsubscript{eq}) for the LFN trials, with a high of 64.2dB at 100Hz. The results for the 80Hz and 100Hz trials, taking the TOB centers of 80Hz and 100Hz, produce A-weighted levels exceeding those proposed in the reference curve.
The smart speaker produced a mean average of 59.37dB for the LFN trials, with a high of 71.6dB at 100Hz. All three tests for this device exceed the corresponding levels proposed in the curve for the TOB center frequencies.
Finally, the headphones produced a mean average of 39.2dB for the LFN trials, with a high of 40.2dB at 100Hz. Taking a TOB center frequency of 100Hz, this exceeds the corresponding level in the reference curve.

\begin{table}[t]
\small
\centering
\resizebox{0.9\linewidth}{!}{
\begin{tabular}{@{}lrr@{}}
\toprule
 & \textbf{TOB Center (Hz)} & \textbf{Level} \\ \midrule
Smart speaker (60Hz) & 200 & 64.2 \\
Smart speaker (80Hz) & 160 & 72.5 \\
Smart speaker (100Hz) & 200 & 73.5 \\
Smart speaker (17kHz) & 6,300 & 75.1 \\
Headphones 100Hz & 125 & 39.5 \\
Headphones (17kHz) & 12,500 & 44.2 \\
Headphones (19kHz) & 1,000 & 23.6 \\
Headphones (21kHz) & 1,250 & 23.9 \\
Loudspeaker (80Hz) & 250 & 65.6 \\
Loudspeaker (100Hz) & 500 & 69.0 \\
Vehicle PA (17kHz) & 1,600 & 60.8 \\
Parametric speaker (17kHz) & 12,500 & 74.3 \\
Parametric speaker (19kHz) & 12,500 & 71.2 \\
Parametric speaker (21kHz) & 12,500 & 69.4 \\
Parametric speaker (40kHz) & 12,500 & 75.2 \\ \bottomrule
\end{tabular}
}
\vspace{-0.1cm}
\caption{Components outside our tested ranges, observed during LFN and HFN trials, between 125Hz and 12500Hz TOB centers. Levels shown in LZ\textsubscript{eq}.}
\label{tab:audible}
\vspace{-0.4cm}
\end{table}

\subsection{Audible Components}
\label{sec:resultsaudible}
In some cases, we observed that additional components outside our tested ranges, and therefore more likely to be audible, were also generated at significant levels. The highest levels, i.e., between 125Hz and 12500Hz (TOB centers), are reported in Table~\ref{tab:audible} for each device tested.

Note that sounds at other frequencies may not always present a significant obstacle to an attacker. The headphones, for instance, produced relatively low noise at other frequencies, which would likely go unnoticed. %
However, other devices produced substantial noise at other frequencies. The parametric speaker in particular produced sound of relatively high levels at 12.5kHz. %
Therefore, these issues may present significant obstacles to an attacker wishing to remain covert, depending on variables such as ambient noise, the environment, and the ability of users to perceive sounds at certain frequencies.

\subsection{Temperature and Damage}
\label{sec:temp}
As mentioned earlier, we also took temperature calculations, which we report in Table~\ref{tab:temp}.
These are calculated by subtracting the post-trial measurement from the pre-trial one, in degrees Celsius, and reveal significant temperature increases in the vehicle-mounted PA system during the emission of LFN, and some smaller increases with both the parametric and smart speakers during both the LFN and HFN tests.

\begin{table}[t]
\setlength{\tabcolsep}{3pt}
\centering
\small
\resizebox{\linewidth}{!}{
\begin{tabular}{@{}lrrrrrrrr@{}}
\toprule
 & \multicolumn{4}{c}{\textbf{HFN (kHz)}} & \multicolumn{3}{c}{\textbf{LFN (Hz)}} \\ \midrule
 & {\bf 17kHz} & {\bf 19kHz} & {\bf 21kHz} & {\bf 21kHz} & {\bf 60Hz} & {\bf 80Hz} & {\bf 100Hz} \\
 & & & & {\bf (HPE)}\\ \midrule
Laptop & 0.8 & 0.1 & 0.9 & 1 & 0.8 & 0 & 0.1 \\
Mobile phone & 0.7 & 1.1 & -1.7 & 0 & 1.1 & 0.1 & -0.1 \\
Loudspeaker\hspace{-0.3cm} & 1.6 & 0.6 & 0 & 0.1 & 0.8 & 2.6 & 0.8 \\
Smart speaker & 4.8 & -0.8 & -0.3 & 0.2 & 9.2 & -1.7 & 0.8 \\
Headphones & 0.2 & 0.4 & 1 & 0 & 0.8 & 0.5 & 0.2 \\
3 Laptops & 2 & 0.4 & 0 & 0.2 & 2.4 & 1.3 & 0 \\
3 Phones & 0 & 1 & 0 & 0.7 & 0.9 & 2.3 & 0.1 \\
Vehicle PA & -0.1 & 0.2 & -0.4 & 0.2 & 20.9 & 3.7 & 8.2 \\
Parametric speaker\hspace{-0.5cm} & 4.7 & 4.2 & 3.8 & 8 & 3.3 & 0.6 & 0.4 \\
Vibration speaker\hspace{-0.3cm} & 1.3 & 0.4 & 0.5 & 0.1 & 0.7 & 2.8 & 2 \\ \bottomrule
\end{tabular}
}
\vspace{-0.15cm}
\caption{Temperature changes in degrees Celsius following readings taken before and after each ten-minute trial.}
\label{tab:temp}
\vspace{-0.2cm}
\end{table}

\begin{figure*}[htbp]
\centering
\begin{subfigure}{0.75\textwidth}
  \includegraphics[{width=\linewidth}]{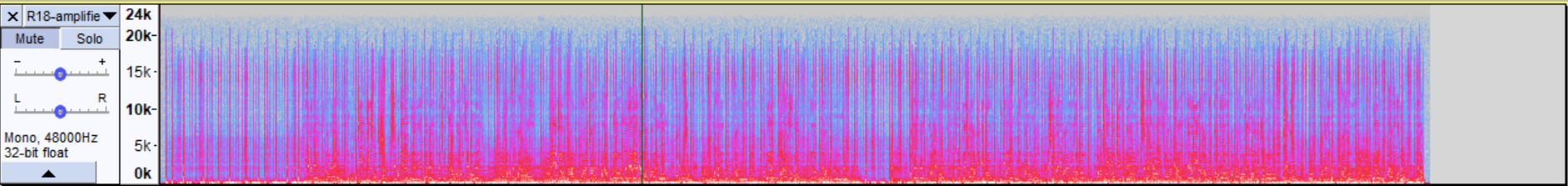}
  \caption{Pre-Experiment}
\end{subfigure}\\
\begin{subfigure}{0.75\textwidth}
  \includegraphics[{width=\linewidth}]{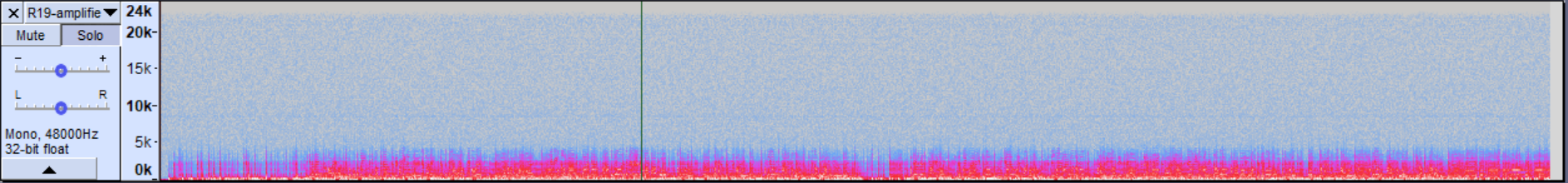}
  \caption{Post-Experiment}
\end{subfigure}
\vspace{-0.2cm}
  \caption{Spectrograms in the audio editor software Audacity, for the pre-experiment recording (a) and post-experiment recording (b) of the smart speaker.}
  \label{fig:spectrograms}
\end{figure*}

We could not replicate the significant temperature increases observed in the smart speaker during the pilot study. 
However, we did note a similar burning odor following the HFN test runs for the smart speaker, and observed a similar degradation in performance. Examining the time history from the sound level meter logs allowed us to investigate this further, and we noted that the speaker appeared to have experienced a marked and critical decrease in performance after approximately five minutes of emitting a 17kHz tone at maximum volume, from which the speaker did not recover. %

To assess the damage to the smart speaker, we made two recordings of an audio track -- a piece of popular music -- played through a {\em newly} purchased smart speaker, in the anechoic chamber. One recording was made before testing, and the other after the HFN test runs had been completed. Comparing the recordings, we observed a significant decrease in the quality of the sound. Further examination using spectrograms, shown in Fig. ~\ref{fig:spectrograms}, show that the speaker appeared to have lost the ability to reproduce frequencies above approximately 5kHz.
This effect, which may be the result of some sort of internal overheating or similar damage, appears to be permanent.

\subsection{Disclosure}
\label{sec:disclosure}
Following our discovery that playing HFN at maximum volume appeared to have caused permanent damage to the smart speaker, affecting the speaker's ability to reproduce frequencies higher than approximately 5kHz, we informed the manufacturer in the spirit of responsible disclosure. 
We found the manufacturer to be responsive and, upon their request, provided them with additional data in order to replicate the issue. 
We received notification approximately two months after initial disclosure that an update would be rolled out in early 2019 to resolve the problem, however, at the time of writing, this has not yet been confirmed.

Also note that we are not aware of any mitigation being applied to address the known issue of unauthenticated control, be it on a local network or remotely if a speaker is exposed on the Internet.

We have not disclosed issues relating to the emission of HFN or LFN for any of the other devices as these are not addressable vulnerabilities as such.
Rather, our attacks demonstrate repurposing of intended functionality.

\section{Discussion}
\label{sec:discussion}
We now provide a broader discussion of our work and its implications.

\subsection{Results}
Out of the ten device set-ups we tested in our experiments, we found that four (two smart, two traditional) were capable of emitting HFN and/or LFN at levels exceeding the averages of those deemed permissible by various bodies such as those referenced in Table~\ref{tab:leighton_table}. More precisely:
\begin{enumerate}
    \item The smart speaker and the headphones exceeded levels for both HFN and LFN; 
    \item The parametric speaker for HFN; 
    \item The loudspeaker for LFN.
\end{enumerate}
In particular, the headphones produced HFN at high levels across all three tested frequencies, and LFN at 100Hz. The smart speaker also produced HFN at high levels at 17kHz, and LFN for all three tested frequencies.

Both of these devices could prove attractive to attackers seeking to attack smart devices in order to produce acoustic effects. 
Indeed, headphones are being increasingly used in developed countries~\cite{henderson2011}, often at high volumes and associated with decreased hearing acuity and hearing loss~\cite{vaidya2018evaluation,mohammadpoorasl2018prevalence}, particularly among young people~\cite{vogel2007,herrera2016}.
Moreover, as mentioned previously, they are often connected to devices such as laptops, mobile phones, and tablets.
Therefore, an attacker could use malware payloads or remote attacks like those presented in this paper to try and cause direct harm to headphone users. 

There are also a number of other attacks, such as the browser or audio manipulation techniques described in Section~\ref{sec:additional_attacks}, which could be used to target such users. 
A variation of the laptop or phone attacks (presented in Section~\ref{sec:attacks}) could also be used to trigger the delivery of sound only when the malware detects that headphones are attached. %

It is also possible that the smart speaker is capable of producing HFN at high levels, and our results indicate that this appears the case for a short period of time. However, this led to the speaker suffering permanent damage (see Section~\ref{sec:temp}), thereby presenting a separate risk: as these speakers do not require authentication for control, an attacker on the local network, or able to access a speaker remotely over the Internet, can cause permanent damage to the device. This is a new class of denial-of-service attack which we had not initially considered, namely, the ability to remotely degrade or damage equipment as a consequence of playing sounds in particular frequency ranges at high volume. In our pilot study, we also found the smart speaker to get hot, but could not replicate the issue; nonetheless, we leave it to future work to assess whether an attacker could cause equipment to overheat, potentially presenting fire hazards to users.
That being said, we hope that the mitigation promised by the manufacturer will address these issues. 
It remains to be seen whether high HFN and LFN levels can be produced after this mitigation has been applied, and whether other speakers are also vulnerable in this respect.
Conversely, the parametric speaker, as a lightweight and portable standalone device with no remote attack surface, may be more suitable as a standalone low-cost HFN acoustic weapon, particularly given its directional capability.
In any case, the use of this device in a public context may present a health risk. 

The loudspeaker was also capable of producing LFN at levels in excess of those associated with annoyance, at two frequencies. However, as a more traditional device, exploitation of this speaker would present some challenges, and the attacker would need to be both within a certain range and able to pair to the speaker.

\subsection{Attack Scenarios}
Overall, there are several scenarios where an attacker may wish to deploy attacks like the ones empirically demonstrated in our experiments. 
The physical and subjective psychological effects associated with long-term exposure to high levels of high or low frequency sound may be of interest to attackers wishing to negatively affect, for instance, the performance of a particular employee, or an organization at scale using multiple infections and attacks. 

If successful, the result of multiple people suffering from noise-induced hearing loss, whether temporary or permanent, in addition to possible negative psychological effects, could have negative ramifications for an organization, vis-\`a-vis their productivity or profitability. It would thus be possible for an attack to not only rely on, e.g., espionage, data theft, or sabotage, but also on acoustic attacks. It may also be possible for these attacks to be used as part of a targeted harassment campaign on an individual basis.

Additionally, as discussed in Section~\ref{sec:relatedwork}, there are potential other uses of HFN for attacks, including covert exfiltration of data; ultrasonic monitoring; and corruption of disk drives. Depending on the attacker's motivations, these may be other attractive solutions not necessarily constrained by human imperceptibility of, or susceptibility to, HFN, or by MPSPLs.

\subsection{Attack Feasibility}
As the number of connected, audio-capable devices continues to rise as a result of the proliferation of smart IoT devices and an increasingly networked society, it is likely that the number of devices vulnerable to these attacks will also rise. 
Therefore, the attack surface for this class of attacks may also grow considerably, and cases in which adversaries view such attacks as a viable option may also increase.

The attacks and malware we developed for this study, while only proof-of-concept and decidedly non-covert relative to most ``in-the wild'' malware, were relatively easy to create and deploy. 
It was not particularly difficult, for instance, to develop malware that could be used to manipulate system volume, or to attack a smart speaker and cause it to stream audio of our choosing. 
Therefore, we argue that our work only scratches the surface of what malware can achieve in terms of using connected devices for acoustic attacks, and that, given the ease and scale at which they could be deployed, the attacks discussed in this paper should be of concern.

\subsection{Limitations}
Naturally, our work is not without limitations.
Our experiments were conducted on a relatively small scale and with a limited number of devices, as we aimed to provide a feasibility study of an understudied problem. 
Moreover, due to constraints on the availability of the anechoic chamber, we limited our testing to short exposure times of ten minutes per frequency per device. 
We hope that future research in this area will examine the effects of these attacks on equipment over longer periods -- as the consistent emission of HFN or LFN at high volumes may significantly degrade electronic components, rendering these attacks much less effective.
Moreover, to a large extent, successful acoustic attacks need to rely on (a) the attacker being able to manipulate a given device to emit sufficient levels of noise; (b) the victim not perceiving the emitted audio; (c) the victim being susceptible to the effects; and (d) the device being capable of producing high levels over time. While we have empirically demonstrated (a), and (d) to a certain extent, we acknowledge that further experiments would be required with respect to (b) and (c) especially. 
However, we are obviously constrained in carrying out these experiments by ethics regarding human experimentation and the safety of study participants.

Previous research has examined the effects of HFN and LFN on humans, albeit at attenuated levels~\cite{fletcher2018part1}, which has allowed us to extrapolate findings to real-world effects; this remains a limitation both in terms of assessing actual effects and in determining if the tones deployed, or artifacts thereof, would be perceived. As discussed above, some (but not all) of the tested devices emitted noise at frequencies and levels more likely to be perceived, which could therefore compromise the covert nature of the attacks.

\section{Countermeasures}
\label{sec:countermeasures}
We now discuss possible avenues to mitigate the acoustic attacks presented in this paper.
Specifically, we consider specific countermeasures besides generic ones like restricting the installation or the execution of unauthorized code. %
One avenue would be to follow suggestions by Deshotels~\cite{deshotels2014} about prevention/detection of imperceptible sound as a covert channel. These include limiting the frequency range of speakers to frequencies in the typically audible range; visibly alerting users when device speakers are in use; filtering files during processing, such that frequencies outside the audible range are removed; and, in the case of mobile devices, implementing a permissions restriction on the use of speakers by apps, so that a user has to manually approve this.

As a proof-of-concept, we adapted an existing open-source software project\footnote{\url{https://www.codeproject.com/Articles/22951/Sound-Activated-Recorder-with-Spectrogram-in-C}}, originally intended to be a sound-activated recorder and audio visualisation tool for Windows, to show alerts when noise above certain frequency ranges and user-specified thresholds is detected.
Source code for this application is available on request. 
Naturally, this approach does rely on the capabilities of the microphone and soundcard on the host, arguably making it somewhat unrealistic for everyday consumer use, particularly in the case of true ultrasonic sound, or lower frequency sound sub-50Hz. %

A similar approach could be used for mobile-based detection. In our pilot study, we used two free Android apps from the Google Play Store, along with a relatively inexpensive external microphone, and found that they were able to generate alerts when sounds exceeded certain levels, particularly with HFN. A wide range of other apps, for both iOS and Android, may be suitable for noise measurements, as a low-cost alternative to traditional SLMs~\cite{kardous2014,murphy2016testing,roberts2016improving}, albeit within device limitations and with the caveat that there may be a decrease in accuracy. While many of these apps do not target HFN or LFN specifically, they may be able to generate alerts when certain level thresholds are exceeded.

It also remains crucial that employers comply with applicable legislation pertaining to acceptable noise limits. %
As noted in Section~\ref{sec:relatedwork}, while a number of guidelines and measurement and assessment criteria exist for both LFN and HFN, researchers have argued that these may be inadequate due to methodological issues~\cite{leighton2016}, underestimation of effects~\cite{leventhall2003,vasudevan1982}, and a lack of clarity on the applicability of occupational guidelines in other contexts, such as public exposure~\cite{leighton2018}.

An additional countermeasure could be to include heuristic features in consumer and enterprise antivirus detection engines, aiming to detect these attacks. For instance, certain behaviors, taken in combination, are often detected by antivirus engines as suspicious, and the user may be presented with a confirmation prompt to indicate whether or not this is expected behavior~\cite{koret2015}. %
In this context, there are few legitimate reasons for applications to need to alter the system or media volume.

As mentioned in Section~\ref{sec:disclosure}, we were informed that the issue affecting the smart speaker, which enabled an attacker on the same network or with remote access to the speaker (e.g., over the Internet, via DNS rebinding~\cite{dns}) to cause permanent damage to the device by playing HFN at high volumes, would be resolved by the vendor, but at the time of writing this has not been confirmed. 
In any case, we advise users owning smart speakers that allow control of certain functions (playing/streaming audio, changing volume) over a network to not employ port forwarding or UPnP, which would expose their speaker to potential remote attack. 
Where the control of such speakers over an API remains unauthenticated, this may still present a risk on a local network. 

Finally, we argue that effective countermeasures mitigating the attacks presented in this paper could also be deployed to detect covert transmissions using ultrasonic audio, an active area of research as applied to both ultrasonic tracking, with subsequent privacy applications~\cite{mavroudis2017}, and to air-gap bypasses~\cite{hanspach2014,deshotels2014,wixey2017}.
\section{Conclusion}
\label{sec:conclusion}
This paper presented a novel class of attack, combining existing and new proof-of-concept malware and attacks to cause ordinary consumer devices to produce high-frequency noise (HFN) and low-frequency noise (LFN) at high levels. 
We empirically verified these attacks on a number of commodity hardware devices.
Specifically, we found that a few devices appear to be capable of producing potentially imperceptible sounds at levels at or exceeding several recommended thresholds, as a direct result.

Like other researchers who previously attempted to examine the psychological and physical effects of high and low frequencies on humans, we found that the lack of consensus for established and adequate safety guidelines for HFN and LFN frequencies represents a challenge toward assessing the real-world consequences of these attacks. 

As societies become more reliant on networks and connectivity, and as digital and physical worlds become more integrated, we believe that attackers may become increasingly interested in leveraging digital vulnerabilities against human users. In this paper, we found that the triviality of executing these attacks, and the size of the potential attack surface, could mean that the repurposing of consumer equipment for acoustic attacks may be viable for attackers aiming to directly cause harm to humans. 

In future work, we plan to examine the capabilities of a wider range of equipment, in a variety of environments and at different distances. In particular, testing other smart speakers and headphones will provide a better understanding of the threats these devices may present. Moreover, for practical reasons, we limited our research to an assessment of consumer products which were relatively inexpensive and portable, and took measurements in an anechoic chamber at a distance of one meter. However, our attacks could be applied to larger and more powerful equipment with the potential to affect many more people in a wider area and to a much greater extent. For instance, an attack against a connected PA system at a music or sporting event, or against the speaker system in a vehicle, could produce audio at much more harmful levels. Other, more ``noisy'' channels, such as smart television broadcasts, or injecting HFN or LFN into phone conversations, may also be effective, particularly as the presence of other, more audible frequencies in such channels may decrease the likelihood of HFN/LFN being perceived by the victim.

We will also examine the applicability of these attacks to offensive cyber-campaigns at scale. For instance, an attack against an organization whereby many co-located user laptops in an office environment are infected with a self-replicating worm, using a payload similar to our proof-of-concept Windows malware, could result in users being exposed to more harmful levels of audio, for longer durations.

\descr{Availability.} As mentioned earlier, we have not released the code of our proof-of-concept attacks, nor the specifications of the devices in our experiments, in order to minimize the risk to the general public.

\small
\bibliographystyle{abbrv}
\bibliography{references}

\begin{thebibliography}{10}

\bibitem{acton1967}
W.~Acton and M.~Carson.
\newblock {Auditory and subjective effects of airborne noise from industrial
  ultrasonic sources}.
\newblock {\em Occupational and Environmental Medicine}, 24(4), 1967.

\bibitem{altmann2001}
J.~Altmann.
\newblock {Acoustic weapons-a prospective assessment}.
\newblock {\em Science \& Global Security}, 9(3), 2001.

\bibitem{andringa2013}
T.~C. Andringa and J.~J.~L. Lanser.
\newblock {How pleasant sounds promote and annoying sounds impede health: A
  cognitive approach}.
\newblock {\em International journal of environmental research and public
  health}, 10(4), 2013.

\bibitem{arkin1997}
W.~M. Arkin.
\newblock {Acoustic anti-personnel weapons: An inhumane future?}
\newblock {\em Medicine, Conflict and Survival}, 13(4), 1997.

\bibitem{ashihara2006}
K.~Ashihara, K.~Kurakata, T.~Mizunami, and K.~Matsushita.
\newblock {Hearing threshold for pure tones above 20 kHz}.
\newblock {\em Acoustical science and technology}, 27(1), 2006.

\bibitem{bartholomew2018}
R.~E. Bartholomew and D.~F.~Z. P{\'e}rez.
\newblock {Chasing ghosts in Cuba: Is mass psychogenic illness masquerading as
  an acoustical attack?}
\newblock {\em International Journal of Social Psychiatry}, 64(5), 2018.

\bibitem{ben2013towards}
L.~Ben~Othmane, A.~Al-Fuqaha, E.~ben Hamida, and M.~Van Den~Brand.
\newblock {Towards extended safety in connected vehicles}.
\newblock In {\em {16th International IEEE Conference on Intelligent
  Transportation Systems}}, 2013.

\bibitem{bengtsson2003}
J.~Bengtsson.
\newblock {\em {Low Frequency Noise During Work: Effects on Performance and
  Annoyance}}.
\newblock PhD thesis, G{\"o}tenburg University, 2003.

\bibitem{benignus1975}
V.~A. Benignus, D.~A. Otto, and J.~H. Knelson.
\newblock {Effect of low-frequency random noises on performance of a numeric
  monitoring task}.
\newblock {\em Perceptual and motor skills}, 40(1), 1975.

\bibitem{bolin2011}
K.~Bolin, G.~Bluhm, G.~Eriksson, and M.~E. Nilsson.
\newblock {Infrasound and low frequency noise from wind turbines: exposure and
  health effects}.
\newblock {\em Environmental research letters}, 6(3), 2011.

\bibitem{bolton2018}
C.~Bolton, S.~Rampazzi, C.~Li, A.~Kwong, W.~Xu, and K.~Fu.
\newblock {Blue Note: How Intentional Acoustic Interference Damages
  Availability and Integrity in Hard Disk Drives and Operating Systems}.
\newblock In {\em {IEEE Symposium on Security \& Privacy}}, 2018.

\bibitem{brody2018malware}
R.~G. Brody, H.~U. Chang, and E.~S. Schoenberg.
\newblock {Malware at its worst: death and destruction}.
\newblock {\em International Journal of Accounting \& Information Management},
  26(4), 2018.

\bibitem{chen2014}
T.~M. Chen.
\newblock {Cyberterrorism after Stuxnet}.
\newblock Technical report, Strategic Studies Institute, Army War College
  (U.S.), 2014.

\bibitem{chopra2016}
A.~Chopra, B.~S. Thomas, K.~Mohan, and K.~Sivaraman.
\newblock {Auditory and Nonauditory Effects of Ultrasonic Scaler Use and Its
  Role in the Development of Permanent Hearing Loss}.
\newblock {\em Oral Health \& Preventive Dentistry}, 14(6), 2016.

\bibitem{cocchi1992}
A.~Cocchi, P.~Fausti, and S.~Piva.
\newblock {Experimental characterisation of the low frequency noise annoyance
  arising from industrial plants}.
\newblock {\em Journal of Low Frequency Noise, Vibration and Active Control},
  11(4), 1992.

\bibitem{cunche2018}
M.~Cunche and L.~S. Cardoso.
\newblock {Analyzing Ultrasound-based Physical Tracking Systems}.
\newblock In {\em {GreHack}}, 2018.

\bibitem{deshotels2014}
L.~Deshotels.
\newblock {Inaudible Sound as a Covert Channel in Mobile Devices}.
\newblock In {\em {WOOT}}, 2014.

\bibitem{dolder2018}
C.~N. Dolder, M.~D. Fletcher, S.~Lloyd~Jones, B.~Lineton, S.~R. Dennison,
  M.~Symmonds, P.~R. White, and T.~G. Leighton.
\newblock {Measurements of ultrasonic deterrents and an acoustically branded
  hairdryer: Ambiguities in guideline compliance}.
\newblock {\em Journal of the Acoustical Society of America}, 144(4), 2018.

\bibitem{duckleighton2018}
F.~Duck and T.~G. Leighton.
\newblock {Frequency bands for ultrasound, suitable for the consideration of
  its health effects}.
\newblock {\em Journal of the Acoustical Society of America}, 144(4), 2018.

\bibitem{durrant1995}
J.~D. Durrant and J.~H. Lovrinic.
\newblock {\em {Bases of hearing science}}.
\newblock Lippincott Williams and Wilkins, 1995.

\bibitem{filonenko2010}
V.~Filonenko, C.~Cullen, and J.~Carswell.
\newblock {Investigating Ultrasonic Positioning on Mobile Phones}.
\newblock In {\em International Conference on Indoor Positioning and Indoor
  Navigation}, 2010.

\bibitem{fletcher2018part1}
M.~D. Fletcher, S.~Lloyd~Jones, P.~R. White, C.~N. Dolder, T.~G. Leighton, and
  B.~Lineton.
\newblock {Effects of very high-frequency sound and ultrasound on humans. Part
  I: Adverse symptoms after exposure to audible very-high frequency sound}.
\newblock {\em Journal of the Acoustical Society of America}, 144(4), 2018.

\bibitem{fletcher2018effects2}
M.~D. Fletcher, S.~Lloyd~Jones, P.~R. White, C.~N. Dolder, T.~G. Leighton, and
  B.~Lineton.
\newblock {Effects of very high-frequency sound and ultrasound on humans. Part
  II: A double-blind randomized provocation study of inaudible 20-kHz
  ultrasound}.
\newblock {\em Journal of the Acoustical Society of America}, 144(4), 2018.

\bibitem{grzesik1986}
J.~Grzesik and E.~Pluta.
\newblock {Dynamics of high-frequency hearing loss of operators of industrial
  ultrasonic devices}.
\newblock {\em International archives of occupational and environmental
  health}, 57(2), 1986.

\bibitem{infradetector}
S.~Gudkov.
\newblock {Google Play Store -- InfraSound Detector App}.
\newblock
  \url{https://play.google.com/store/apps/details?id=com.microcadsystems.serge.infrasounddetector},
  2019.

\bibitem{ultradetector}
S.~Gudkov.
\newblock {Google Play Store -- UltraSound Detector App}.
\newblock
  \url{https://play.google.com/store/apps/details?id=com.microcadsystems.serge.ultrasounddetector},
  2019.

\bibitem{halperin2008pacemakers}
D.~Halperin, T.~S. Heydt-Benjamin, B.~Ransford, S.~S. Clark, B.~Defend,
  W.~Morgan, K.~Fu, T.~Kohno, and W.~H. Maisel.
\newblock {Pacemakers and implantable cardiac defibrillators: Software radio
  attacks and zero-power defenses}.
\newblock In {\em {IEEE Symposium on Security \& Privacy}}, 2008.

\bibitem{hansen2001fundamentals}
C.~H. Hansen.
\newblock {Fundamentals of acoustics}.
\newblock {\em Occupational Exposure to Noise: Evaluation, Prevention and
  Control. World Health Organization}, 2001.

\bibitem{hanspach2014}
M.~Hanspach and M.~Goetz.
\newblock {On covert acoustical mesh networks in air}.
\newblock {\em arXiv preprint 1406.1213}, 2014.

\bibitem{dns}
L.~Hay~Newman.
\newblock {Millions of streaming devices are vulnerable to a retro Web attack}.
\newblock
  \url{https://www.wired.com/story/chromecast-roku-sonos-dns-rebinding-vulnerability/},
  2018.

\bibitem{henderson2011}
E.~Henderson, M.~A. Testa, and C.~Hartnick.
\newblock {Prevalence of noise-induced hearing-threshold shifts and hearing
  loss among US youths}.
\newblock {\em Pediatrics}, 127(1), 2011.

\bibitem{herrera2016}
S.~Herrera, A.~B.~M. de~Lacerda, D.~Lurdes, P.~A. Alcaras, L.~H. Ribeiro,
  et~al.
\newblock {Amplified music with headphones and its implications on hearing
  health in teens}.
\newblock {\em International Tinnitus Journal}, 20(1), 2016.

\bibitem{howard2005}
C.~Q. Howard, C.~H. Hansen, and A.~C. Zander.
\newblock {A review of current ultrasound exposure limits}.
\newblock {\em Journal of Occupational Health and Safety of Australia and New
  Zealand}, 21(3), 2005.

\bibitem{kaczmarska2007}
A.~Kaczmarska and A.~{\L}uczak.
\newblock {A study of annoyance caused by low-frequency noise during mental
  work}.
\newblock {\em International Journal of Occupational Safety and Ergonomics},
  13(2), 2007.

\bibitem{kardous2014}
C.~A. Kardous and P.~B. Shaw.
\newblock {Evaluation of smartphone sound measurement applications}.
\newblock {\em Journal of the Acoustical Society of America}, 135(4), 2014.

\bibitem{kjellberg1984}
A.~Kjellberg, M.~Goldstein, and F.~Gamberale.
\newblock {An assesment of dB (A) for predicting loudness and annoyance of
  noise containing low frequency components}.
\newblock {\em Journal of Low Frequency Noise, Vibration and Active Control},
  3(3), 1984.

\bibitem{koch2017}
C.~Koch.
\newblock {Hearing beyond the limit: Measurement, perception and impact of
  infrasound and ultrasonic noise}.
\newblock In {\em {12th ICBEN Congress on Noise as a Public Health Problem}},
  2017.

\bibitem{kolias2017ddos}
C.~Kolias, G.~Kambourakis, A.~Stavrou, and J.~Voas.
\newblock {DDoS in the IoT: Mirai and Other Botnets}.
\newblock {\em IEEE Computer}, 50(7), 2017.

\bibitem{koret2015}
J.~Koret and E.~Bachaalany.
\newblock {\em {The antivirus hacker's handbook}}.
\newblock John Wiley \& Sons, 2015.

\bibitem{krotofil2014}
M.~Krotofil.
\newblock {Cyber Can Kill and Destroy Too: Blurring Borders Between
  Conventional and Cyber Warfare}.
\newblock {\em International Journal of Cyber Warfare and Terrorism (IJCWT)},
  4(3), 2014.

\bibitem{lawton2001}
B.~W. Lawton.
\newblock {Damage to human hearing by airborne sound of very high frequency or
  ultrasonic frequency}.
\newblock \url{http://www.hse.gov.uk/research/crr_pdf/2001/crr01343.pdf}, 2001.

\bibitem{leighton2016}
T.~G. Leighton.
\newblock {Are some people suffering as a result of increasing mass exposure of
  the public to ultrasound in air?}
\newblock {\em Proceedings of the Royal Society A: Mathematical, Physical and
  Engineering Sciences}, 472(2185), 2016.

\bibitem{leighton2018}
T.~G. Leighton.
\newblock {Ultrasound in air—Guidelines, applications, public exposures, and
  claims of attacks in Cuba and China}.
\newblock {\em Journal of the Acoustical Society of America}, 144(4), 2018.

\bibitem{leventhall2009}
G.~Leventhall.
\newblock {Low Frequency Noise. What we know, what we do not know, and what we
  would like to know}.
\newblock {\em Journal of Low Frequency Noise, Vibration and Active Control},
  28(2), 2009.

\bibitem{leventhall2003}
G.~Leventhall, P.~Pelmear, and S.~Benton.
\newblock {A review of published research on low frequency noise and its
  effects}.
\newblock
  \url{https://westminsterresearch.westminster.ac.uk/item/935y3/a-review-of-published-research-on-low-frequency-noise-and-its-effects},
  2003.

\bibitem{macca2015}
I.~Macc{\`a}, M.~L. Scapellato, M.~Carrieri, S.~Maso, A.~Trevisan, and G.~B.
  Bartolucci.
\newblock {High-frequency hearing thresholds: effects of age, occupational
  ultrasound and noise exposure}.
\newblock {\em International Archives of Occupational and Environmental
  Health}, 88(2), 2015.

\bibitem{mavroudis2017}
V.~Mavroudis, S.~Hao, Y.~Fratantonio, F.~Maggi, C.~Kruegel, and G.~Vigna.
\newblock {On the privacy and security of the ultrasound ecosystem}.
\newblock {\em Proceedings on Privacy Enhancing Technologies}, 2017(2), 2017.

\bibitem{mirowska2000}
M.~Mirowska and E.~Mroz.
\newblock {Effect of low frequency noise at low levels on human health in light
  of questionnaire investigation}.
\newblock In {\em {InterNoise 2000}}, 2000.

\bibitem{mohammadpoorasl2018prevalence}
A.~Mohammadpoorasl, M.~Hajizadeh, S.~Marin, P.~Heidari, and M.~Ghalenoee.
\newblock Prevalence and pattern of using headphones and its relationship with
  hearing loss among students.
\newblock {\em Health Scope}, 2018.

\bibitem{moller2002}
H.~M{\o}ller and M.~Lydolf.
\newblock {A questionnaire survey of complaints of infrasound and low-frequency
  noise}.
\newblock {\em Journal of Low Frequency Noise, Vibration and Active Control},
  21(2), 2002.

\bibitem{moorhouse2011}
A.~Moorhouse, D.~Waddington, M.~Adams, et~al.
\newblock {Proposed criteria for the assessment of low frequency noise
  disturbance (Revision 1)}.
\newblock
  \url{http://usir.salford.ac.uk/491/1/NANR45-criteria__rev1_23_12_2011_(2).pdf},
  2011.

\bibitem{muhlhans2017}
J.~H. M{\"u}hlhans.
\newblock {Low frequency and infrasound: A critical review of the myths,
  misbeliefs and their relevance to music perception research}.
\newblock {\em Musicae Scientiae}, 21(3), 2017.

\bibitem{murphy2016testing}
E.~Murphy and E.~A. King.
\newblock {Testing the accuracy of smartphones and sound level meter
  applications for measuring environmental noise}.
\newblock {\em Applied Acoustics}, 106, 2016.

\bibitem{oluwafemi2013}
T.~Oluwafemi, T.~Kohno, S.~Gupta, and S.~Patel.
\newblock {Experimental Security Analyses of Non-Networked Compact Fluorescent
  Lamps: A Case Study of Home Automation Security}.
\newblock In {\em {LASER}}, 2013.

\bibitem{oohashi2000}
T.~Oohashi, E.~Nishina, M.~Honda, Y.~Yonekura, Y.~Fuwamoto, N.~Kawai,
  T.~Maekawa, S.~Nakamura, H.~Fukuyama, and H.~Shibasaki.
\newblock {Inaudible high-frequency sounds affect brain activity: hypersonic
  effect}.
\newblock {\em Journal of neurophysiology}, 83(6), 2000.

\bibitem{pawlaczyk2005}
M.~Pawlaczyk-{\l}uszczy{\'n}ska, A.~Dudarewicz, M.~Waszkowska, W.~Szymczak,
  M.~Kamedu{\l}a, and M.~{\'S}liwi{\'n}ska-Kowalska.
\newblock {Does low frequency noise at modarate levels influence human mental
  performance?}
\newblock {\em Journal of Low Frequency Noise, Vibration, and Active Control},
  24(1), 2005.

\bibitem{perssonbjorkman1988}
K.~Persson and M.~Bjorkman.
\newblock {Annoyance due to low frequency noise and the use of the dB (A)
  scale}.
\newblock {\em Journal of Sound and Vibration}, 127(3), 1988.

\bibitem{persson1988}
K.~Persson and R.~Rylander.
\newblock {Disturbance from low-frequency noise in the environment: A survey
  among the local environmental health authorities in Sweden}.
\newblock {\em Journal of Sound and Vibration}, 121(2), 1988.

\bibitem{pompei2002sound}
F.~J. Pompei.
\newblock {\em {Sound from ultrasound: The parametric array as an audible sound
  source}}.
\newblock PhD thesis, Massachusetts Institute of Technology, 2002.

\bibitem{poulsen2008}
K.~Poulsen.
\newblock {Hackers assault epilepsy patients via computer}.
\newblock
  \url{https://www.wired.com/2008/03/hackers-assault-epilepsy-patients-via-computer/},
  2008.

\bibitem{qibai2004}
C.~Y.~H. Qibai and H.~Shi.
\newblock {An investigation on the physiological and psychological effects of
  infrasound on persons}.
\newblock {\em Journal of Low Frequency Noise, Vibration and Active Control},
  23(1), 2004.

\bibitem{rios2017iot}
B.~Rios and J.~Butts.
\newblock {When IoT Attacks: understanding the safety risks associated with
  connected devices}.
\newblock {\em Black Hat USA}, 2017.

\bibitem{roberts2016improving}
B.~Roberts, C.~Kardous, and R.~Neitzel.
\newblock {Improving the accuracy of smart devices to measure noise exposure}.
\newblock {\em Journal of Occupational and Environmental Hygiene}, 13(11),
  2016.

\bibitem{ronen2016}
E.~Ronen and A.~Shamir.
\newblock {Extended functionality attacks on IoT devices: The case of smart
  lights}.
\newblock In {\em {Security and Privacy (EuroS\&P), 2016 IEEE European
  Symposium on}}, 2016.

\bibitem{rushanan2014sok}
M.~Rushanan, A.~D. Rubin, D.~F. Kune, and C.~M. Swanson.
\newblock {Sok: Security and privacy in implantable medical devices and body
  area networks}.
\newblock In {\em {IEEE Symposium on Security \& Privacy}}, 2014.

\bibitem{smagowska2013}
B.~Smagowska and M.~Pawlaczyk-{\L}uszczy{\'n}ska.
\newblock {Effects of ultrasonic noise on the human body--A bibliographic
  review}.
\newblock {\em International Journal of Occupational Safety and Ergonomics},
  19(2), 2013.

\bibitem{stansfeld2015}
S.~Stansfeld and M.~Shipley.
\newblock {Noise sensitivity and future risk of illness and mortality}.
\newblock {\em Science of the Total Environment}, 520, 2015.

\bibitem{storm2009}
R.~Storm.
\newblock {Health risks due to exposure of low-frequency noise}.
\newblock
  \url{http://www.diva-portal.org/smash/get/diva2:273045/FULLTEXT01.pdf}, 2009.

\bibitem{sturm2014}
L.~D. Sturm, C.~B. Williams, J.~A. Camelio, J.~White, and R.~Parker.
\newblock {Cyber-physical vunerabilities in additive manufacturing systems}.
\newblock {\em Context}, 7(8), 2014.

\bibitem{ueda2014}
M.~Ueda, A.~Ota, and H.~Takahashi.
\newblock {Investigation on high-frequency noise in public space—We tried
  noise abatement measures for displeasure people}.
\newblock In {\em {Proceedings of the 7th Forum Acusticum}}, 2014.

\bibitem{vaidya2018evaluation}
L.~Vaidya, N.~Shah, and A.~H. Mistry.
\newblock Evaluation of hearing acuity in young adults using personal listening
  devices with earphones.
\newblock {\em Int J Basic Appl Physiol}, 7(1):89, 2018.

\bibitem{wieringen2018}
A.~van Wieringen and C.~Glorieux.
\newblock {Assessment of short-term exposure to an ultrasonic rodent repellent
  device}.
\newblock {\em Journal of the Acoustical Society of America}, 144(4), 2018.

\bibitem{vasudevan1982}
R.~Vasudevan and H.~Leventhall.
\newblock {A study of annoyance due to low frequency noise in the home}.
\newblock {\em Journal of Low Frequency Noise, Vibration and Active Control},
  1(4), 1982.

\bibitem{vinokur2004}
R.~Vinokur.
\newblock {Acoustic noise as a non-lethal weapon}.
\newblock {\em Sound and Vibration}, 38(10), 2004.

\bibitem{vogel2007}
I.~Vogel, J.~Brug, C.~P. Van~der Ploeg, and H.~Raat.
\newblock {Young people’s exposure to loud music: a summary of the
  literature}.
\newblock {\em American journal of Preventive Medicine}, 33(2), 2007.

\bibitem{vongierke1992}
H.~E. Von~Gierke and C.~W.~a. Nixon.
\newblock {Damage risk criteria for hearing and human body vibration}.
\newblock {\em Noise and vibration control engineering: principles and
  applications}, 1992.

\bibitem{benton1997}
K.~P. Waye, R.~Rylander, S.~Benton, and H.~Leventhall.
\newblock {Effects on performance and work quality due to low frequency
  ventilation noise}.
\newblock {\em Journal of Sound and Vibration}, 205(4), 1997.

\bibitem{williams2015cybersecurity}
P.~A. Williams and A.~J. Woodward.
\newblock {Cybersecurity vulnerabilities in medical devices: a complex
  environment and multifaceted problem}.
\newblock {\em Medical Devices}, 8, 2015.

\bibitem{wilson2002}
J.~D. Wilson, M.~L. Darby, S.~L. Tolle, and J.~C. Sever~Jr.
\newblock {Effects of occupational ultrasonic noise exposure on hearing of
  dental hygienists: a pilot study}.
\newblock {\em Journal of Dental Hygiene}, 76(4), 2002.

\bibitem{wixey2017}
M.~Wixey.
\newblock {See no evil, hear no evil: Hacking invisibly and silently with light
  and sound}.
\newblock In {\em {DEFCON 25}}, 2017.

\bibitem{yan2016}
C.~Yan, W.~Xu, and J.~Liu.
\newblock {Can you trust autonomous vehicles: Contactless attacks against
  sensors of self-driving vehicle}.
\newblock In {\em {DEFCON 24}}, 2016.

\bibitem{yang2017survey}
Y.~Yang, L.~Wu, G.~Yin, L.~Li, and H.~Zhao.
\newblock {A survey on security and privacy issues in internet-of-things}.
\newblock {\em IEEE Internet of Things Journal}, 4(5), 2017.

\end{thebibliography}

\end{document}